# Algebraic Structure of Central Molecular Chirality Starting from Fischer Projections

SALVATORE CAPOZZIELLO[a]* AND ALESSANDRA LATTANZI[b]*

[a]*Dipartimento di Fisica "E. R. Caianiello", INFN sez. di Napoli* and [b]*Dipartimento di Chimica,
Università di Salerno, Via S. Allende, 84081, Baronissi, Salerno, Italy*
E-mail: lattanzi@unisa.it

*ABSTRACT*   The construction of algebraic structure of central molecular chirality is provided starting from the empirical Fischer projections for tetrahedrons. A matrix representation is given and the algebra of $O(4)$ orthogonal group for rotations and inversions is identified. The result can be generalized to chains of connected tetrahedrons.

From the molecular point of view, chirality embraces all chemical disciplines: organic, biological, inorganic, organometallic chemistry. Asymmetry of atomic configurations is a basic feature to establish their chemical, physical and biological properties. When two molecules have identical structural formulae, but they are not superimposable,[1] they are termed *enantiomers*, even called optical antipodes, since they rotate the polarization plane of light in opposite directions. Chiral molecules having *central chirality* contain stereogenic or asymmetric centers.[2] When two molecules with identical constitution, possessing more than one stereogenic center, do not show a mirror-image relation to each other, they are called *diastereoisomers*.

Other forms of chirality are generated when the rotation around double bonds is prevented as in allenes[3] or is limited due to steric hindrance as in biphenyls[4]: in these cases we are dealing with *axial chirality*.

Chirality can be even due to a helical shape of the molecule which can be left- or right handed in orientation.[5]

In the last decades, discrete mathematics and qualitative descriptions of the spatial features of molecules have provided a large development of theoretical stereochemistry.[6] The aim of these studies is related to the achievement of formal approaches to stereochemical issues of classification. Molecular chirality has been studied by algebraic methods based on permutation group theory and group representation theory,[7] providing general insight into chirality properties. It can be asserted, with no doubt, that the most important three dimensional structure in chemistry is the tetrahedron and chiral tetrahedral molecules are not only the most familiar to the organic chemist but largely widespread compounds.

An extremely useful method to represent tetrahedral molecules was reported in 1891 by Emil Fischer,[1a] who proposed the well-known planar projection formulae. When describing a molecule in this representation, some rules have to be followed (Fig. 1): the atoms pointing sideways must project forward in the model, while those pointing up and down in the projection must extend toward the rear. As an example, let us take into account the (*S*)-(+)-lactic acid.

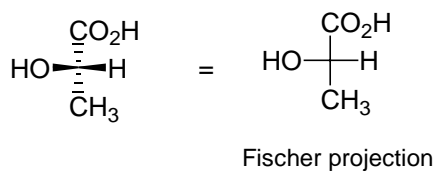

Fischer projection

**Fig. 1.** Fischer projection of (*S*)-(+)-lactic acid

In order to obtain proper results using Fischer projections, they must be treated differently from models in testing superimposability. Projections may not be rotated of 90°, while a 180° rotation is allowed. The interchange of any two groups results in the conversion of an enantiomer into its mirror image (Fig. 2).

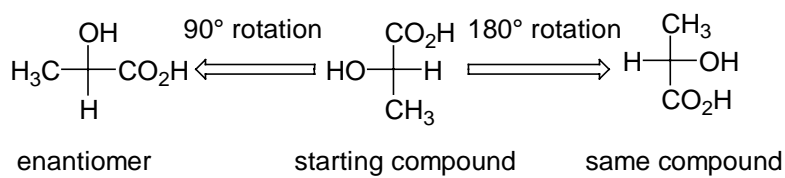

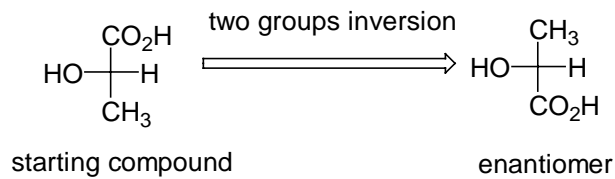

*Fig. 2. Fundamental rules to handle with Fischer projections*

From now on, let us indicate the chemical groups by numbers running from 1 to 4. For the example which we are considering: OH=1, $CO_2H$=2, H=3, $CH_3$=4, without taking into account the effective priorities of the groups.

There are 24 (=4! the number of permutations of 4 ligands among 4 sites) projections. Twelve of these correspond to the (+) enantiomer and are illustrated in Fig. 3.

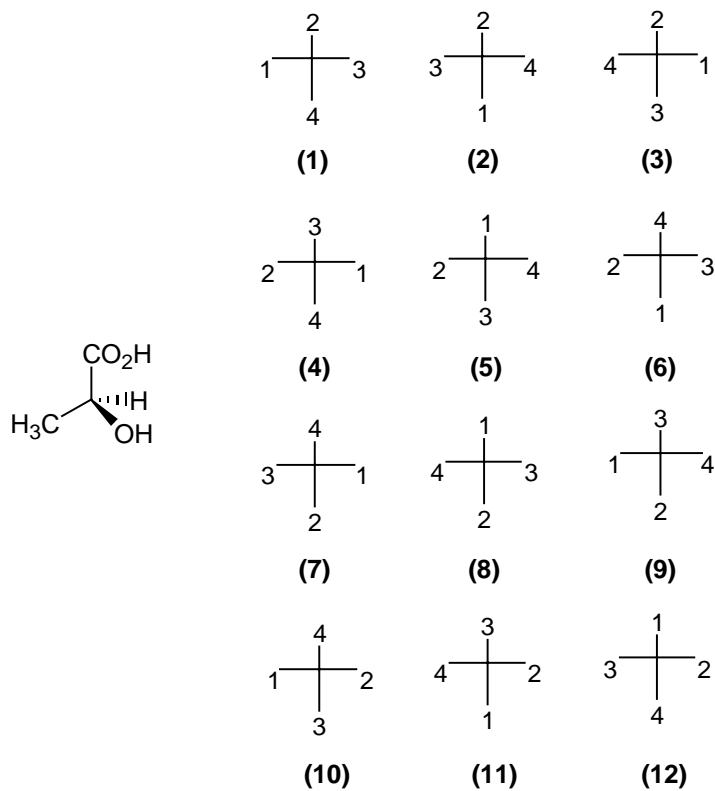

**Fig. 3.** Twelve Fischer projections of (*S*)-(+)-lactic acid The other 12 graphs in Fig. 4 represent the (−) enantiomer.

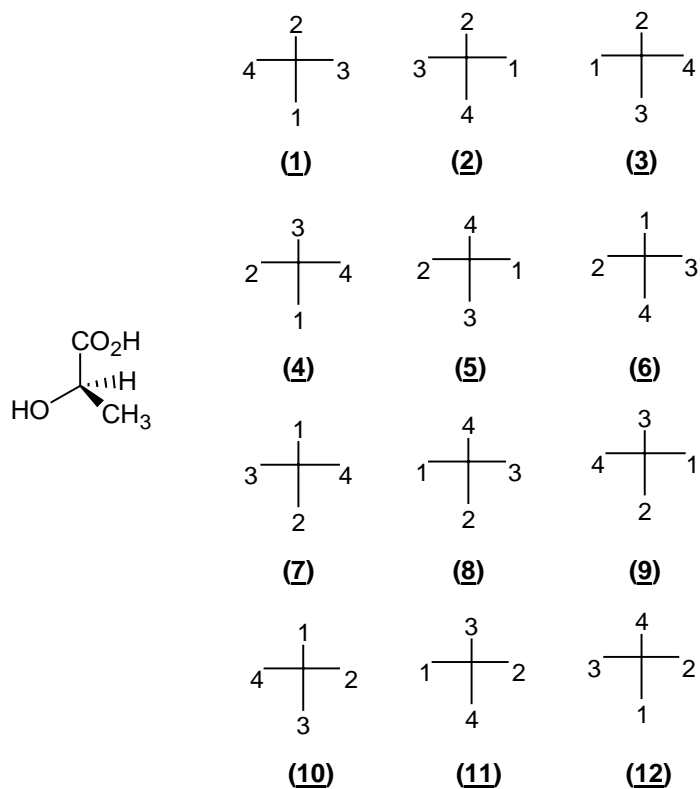

*Fig. 4. Twelve Fischer projections of (R)-(-)-lactic acid*

The permutations shown in Fig. 3 can be obtained, either by permuting groups in threes or by turning the projections by 180°. The permutations drawn in Fig. 4 derive by those in Fig. 3 simply by the interchange of two groups. With these considerations in mind, it is immediate asking for an algebraic structure which can be built from Fischer projections and this is the aim of the present article.

**ALGEBRAIC STRUCTURE OF CENTRAL MOLECULAR CHIRALITY**

In another report,[8] we described a tetrahedral molecule by a projection on a {x,y} plane where the molecule center coincided with the origin of the axes.

The approach, based on complex numbers, is a straightforward way to represent the "length" of the bond with respect to the stereogenic center and the "angular position" with respect to the other bonds. In general, given a tetrahedral molecule with a stereogenic center, we can always project it on a plane containing the stereogenic center.

Every bond, in the plane {x,y}, can be given in polar representation by

$$\Psi_j = \rho_j e^{i\theta_j} \tag{1}$$

where $\rho_j$ is the "modulus", i.e. the projected length of the bond, $\theta_j$ is the "anomaly", i.e. the position of the bond with respect to the x, y axes (and then with respect to the other bonds) having chosen a rotation versus. The number $i = \sqrt{-1}$ is the imaginary unit. A molecule with one stereogenic center is then given by the sum vector

$$\mathcal{M} = \sum_{j=1}^{4} \rho_j e^{i\theta_j} \qquad (2)$$

in any symmetry plane.

In order to reduce the Fischer rules to an algebraic structure, let us define the central chirality operator $\chi_k$ acting on a tetrahedral molecule. From now on, we shall take into account only one tetrahedron, but the generalization of the following results to simply connected chains of tetrahedrons is straight as we shall see below.

A tetrahedral molecule can be assigned by a column vector $\mathcal{M}$, rewriting Eq. (2) as

$$\mathcal{M} = \begin{pmatrix} \Psi_1 \\ \Psi_2 \\ \Psi_3 \\ \Psi_4 \end{pmatrix} \qquad (3)$$

The $\Psi_j$ are defined as in Eq. [1].

The corresponding Fischer projection is

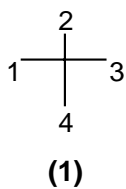

**(1)**

which is the first in Fig. 3. The position of the bonds in the column vector (3) are assigned starting from the left and proceeding clockwise in the Fischer projection.

The matrix representation of the projection **(1)** is assumed as "fundamental", i.e.,

$$\chi_1 = \begin{pmatrix} 1 & 0 & 0 & 0 \\ 0 & 1 & 0 & 0 \\ 0 & 0 & 1 & 0 \\ 0 & 0 & 0 & 1 \end{pmatrix} \tag{4}$$

so the action on the column vector $\mathcal{M}$ is

$$\chi_1 \begin{pmatrix} \Psi_1 \\ \Psi_2 \\ \Psi_3 \\ \Psi_4 \end{pmatrix} = \begin{pmatrix} \Psi_1 \\ \Psi_2 \\ \Psi_3 \\ \Psi_4 \end{pmatrix} \tag{5}$$

$\chi_1$ is nothing else but the identity operator. The configuration **(2)** of Fig. 3 can be achieved as soon as we define an operator $\chi_2$ acting as

$$\chi_2 \begin{pmatrix} \Psi_1 \\ \Psi_2 \\ \Psi_3 \\ \Psi_4 \end{pmatrix} = \begin{pmatrix} \Psi_3 \\ \Psi_2 \\ \Psi_4 \\ \Psi_1 \end{pmatrix} \tag{6}$$

which corresponds to the matrix

$$\chi_2 = \begin{pmatrix} 0 & 0 & 1 & 0 \\ 0 & 1 & 0 & 0 \\ 0 & 0 & 0 & 1 \\ 1 & 0 & 0 & 0 \end{pmatrix} \tag{7}$$

it is clear that $\chi_2$ yields a permutation on the bonds $\Psi_j$. In the specific case, it is a rotation.

On the other hand, the configuration **(1)** of the (−) enantiomer can be obtained starting from the column vector (4), if we define an operator $\bar{\chi}_1$ which acts as

$$\bar{\chi}_1 \begin{pmatrix} \Psi_1 \\ \Psi_2 \\ \Psi_3 \\ \Psi_4 \end{pmatrix} = \begin{pmatrix} \Psi_4 \\ \Psi_2 \\ \Psi_3 \\ \Psi_1 \end{pmatrix} \tag{8}$$

Explicitly, we have

$$\bar{\chi}_1 = \begin{pmatrix} 0 & 0 & 0 & 1 \\ 0 & 1 & 0 & 0 \\ 0 & 0 & 1 & 0 \\ 1 & 0 & 0 & 0 \end{pmatrix} \qquad (9)$$

It generates the inversion between the bonds $\Psi_1$ and $\Psi_4$.

By this approach, we can obtain all the 24 projections (12 for the (+) enantiomer and 12 for the (−) enantiomer represented in Figs. 3 and 4) by matrix operators acting on the fundamental projection (1). The following tables summarize the situation. The operators $\chi_k$ give rise to the representations of the (+) enantiomer, while the operators $\bar{\chi}_k$ give rise to those of the (−) enantiomer. Obviously $k = 1,..,12$.

Table I, (+)-enantiomer:

$$\chi_1 = \begin{pmatrix} 1 & 0 & 0 & 0 \\ 0 & 1 & 0 & 0 \\ 0 & 0 & 1 & 0 \\ 0 & 0 & 0 & 1 \end{pmatrix} \qquad \chi_2 = \begin{pmatrix} 0 & 0 & 1 & 0 \\ 0 & 1 & 0 & 0 \\ 0 & 0 & 0 & 1 \\ 1 & 0 & 0 & 0 \end{pmatrix} \qquad \chi_3 = \begin{pmatrix} 0 & 0 & 0 & 1 \\ 0 & 1 & 0 & 0 \\ 1 & 0 & 0 & 0 \\ 0 & 0 & 1 & 0 \end{pmatrix}$$

$$\chi_4 = \begin{pmatrix} 0 & 1 & 0 & 0 \\ 0 & 0 & 1 & 0 \\ 1 & 0 & 0 & 0 \\ 0 & 0 & 0 & 1 \end{pmatrix} \qquad \chi_5 = \begin{pmatrix} 0 & 1 & 0 & 0 \\ 1 & 0 & 0 & 0 \\ 0 & 0 & 0 & 1 \\ 0 & 0 & 1 & 0 \end{pmatrix} \qquad \chi_6 = \begin{pmatrix} 0 & 1 & 0 & 0 \\ 0 & 0 & 0 & 1 \\ 0 & 0 & 1 & 0 \\ 1 & 0 & 0 & 0 \end{pmatrix}$$

$$\chi_7 = \begin{pmatrix} 0 & 0 & 1 & 0 \\ 0 & 0 & 0 & 1 \\ 1 & 0 & 0 & 0 \\ 0 & 1 & 0 & 0 \end{pmatrix} \qquad \chi_8 = \begin{pmatrix} 0 & 0 & 0 & 1 \\ 1 & 0 & 0 & 0 \\ 0 & 0 & 1 & 0 \\ 0 & 1 & 0 & 0 \end{pmatrix} \qquad \chi_9 = \begin{pmatrix} 1 & 0 & 0 & 0 \\ 0 & 0 & 1 & 0 \\ 0 & 0 & 0 & 1 \\ 0 & 1 & 0 & 0 \end{pmatrix}$$

$$\chi_{10} = \begin{pmatrix} 1 & 0 & 0 & 0 \\ 0 & 0 & 0 & 1 \\ 0 & 1 & 0 & 0 \\ 0 & 0 & 1 & 0 \end{pmatrix} \qquad \chi_{11} = \begin{pmatrix} 0 & 0 & 0 & 1 \\ 0 & 0 & 1 & 0 \\ 0 & 1 & 0 & 0 \\ 1 & 0 & 0 & 0 \end{pmatrix} \qquad \chi_{12} = \begin{pmatrix} 0 & 0 & 1 & 0 \\ 1 & 0 & 0 & 0 \\ 0 & 1 & 0 & 0 \\ 0 & 0 & 0 & 1 \end{pmatrix}$$

Table II, (−) enantiomer:

$$\bar{\chi}_1 = \begin{pmatrix} 0 & 0 & 0 & 1 \\ 0 & 1 & 0 & 0 \\ 0 & 0 & 1 & 0 \\ 1 & 0 & 0 & 0 \end{pmatrix} \quad \bar{\chi}_2 = \begin{pmatrix} 0 & 0 & 1 & 0 \\ 0 & 1 & 0 & 0 \\ 1 & 0 & 0 & 0 \\ 0 & 0 & 0 & 1 \end{pmatrix} \quad \bar{\chi}_3 = \begin{pmatrix} 1 & 0 & 0 & 0 \\ 0 & 1 & 0 & 0 \\ 0 & 0 & 0 & 1 \\ 0 & 0 & 1 & 0 \end{pmatrix}$$

$$\bar{\chi}_4 = \begin{pmatrix} 0 & 1 & 0 & 0 \\ 0 & 0 & 1 & 0 \\ 0 & 0 & 0 & 1 \\ 1 & 0 & 0 & 0 \end{pmatrix} \quad \bar{\chi}_5 = \begin{pmatrix} 0 & 1 & 0 & 0 \\ 0 & 0 & 0 & 1 \\ 1 & 0 & 0 & 0 \\ 0 & 0 & 1 & 0 \end{pmatrix} \quad \bar{\chi}_6 = \begin{pmatrix} 0 & 1 & 0 & 0 \\ 1 & 0 & 0 & 0 \\ 0 & 0 & 1 & 0 \\ 0 & 0 & 0 & 1 \end{pmatrix}$$

$$\bar{\chi}_7 = \begin{pmatrix} 0 & 0 & 1 & 0 \\ 1 & 0 & 0 & 0 \\ 0 & 0 & 0 & 1 \\ 0 & 1 & 0 & 0 \end{pmatrix} \quad \bar{\chi}_8 = \begin{pmatrix} 1 & 0 & 0 & 0 \\ 0 & 0 & 0 & 1 \\ 0 & 0 & 1 & 0 \\ 0 & 1 & 0 & 0 \end{pmatrix} \quad \bar{\chi}_9 = \begin{pmatrix} 0 & 0 & 0 & 1 \\ 0 & 0 & 1 & 0 \\ 1 & 0 & 0 & 0 \\ 0 & 1 & 0 & 0 \end{pmatrix}$$

$$\bar{\chi}_{10} = \begin{pmatrix} 0 & 0 & 0 & 1 \\ 1 & 0 & 0 & 0 \\ 0 & 1 & 0 & 0 \\ 0 & 0 & 1 & 0 \end{pmatrix} \quad \bar{\chi}_{11} = \begin{pmatrix} 1 & 0 & 0 & 0 \\ 0 & 0 & 1 & 0 \\ 0 & 1 & 0 & 0 \\ 0 & 0 & 0 & 1 \end{pmatrix} \quad \bar{\chi}_{12} = \begin{pmatrix} 0 & 0 & 1 & 0 \\ 0 & 0 & 0 & 1 \\ 0 & 1 & 0 & 0 \\ 1 & 0 & 0 & 0 \end{pmatrix}$$

The matrices in Table I and II are the elements of a 4-parameter algebra. Those in Table I are a representation of rotations, while those in table II are inversions. Both sets constitute the group $O(4)$ of 4×4 orthogonal matrices.

The matrices in Table I are the remarkable subgroup $SO(4)$ of 4×4 matrices with determinant +1. The matrices in Table II have determinant −1, being inversions (or reflections). They do not constitute a group since the product of any two of them has determinant +1. This fact means that the product of two inversions generates a rotation (this is obvious by inverting both the couples of bonds in a tetrahedron). In fact, we have

$$\chi_k \chi_l = \chi_m, \qquad \bar{\chi}_k \bar{\chi}_l = \chi_m, \qquad \bar{\chi}_k \chi_l = \bar{\chi}_m \qquad \text{for } k,l,m,=1,...,12 \quad (10)$$

For example, straightforward calculations give

$$\chi_8 \chi_9 = \chi_5, \qquad \bar{\chi}_5 \bar{\chi}_2 = \chi_9, \qquad \bar{\chi}_{10} \chi_{10} = \bar{\chi}_7 \tag{11}$$

and so on. In summary, the product of two rotations is a rotation, the product of two reflections is a rotation, while the product of a reflection and a rotation is again a reflection. In any case, the total algebra is closed and the following commutation relations can be obtained

$$[\chi_k, \chi_l] = 0 \qquad \text{for } k,m = 5,7,11 \qquad (12)$$

$$[\bar{\chi}_k, \bar{\chi}_l] = \chi_m - \chi_n \qquad \text{for } k,l,m,n = 1,...,12 \qquad (13)$$

$$[\chi_k, \chi_l] = \chi_m - \chi_n \qquad \text{for } k,l,m,n = 1,...,12 \text{ with } k,l \neq 5,7,11 \qquad (14)$$

$$[\chi_k, \bar{\chi}_l] = \bar{\chi}_m - \bar{\chi}_n \qquad \text{for } k,l,m,n = 1,...,12 \qquad (15)$$

It is interesting to note that the rotations $\chi_5$, $\chi_7$, $\chi_{11}$ commute among them.

The 24 matrices in Table I and II are not all independent. They can be grouped as different representations of the same operators. To this aim, we make use of a fundamental theorem of algebra which states that all matrices, representing the same operator, have the same characteristic polynomial.[10] In other words, the characteristic equation of a matrix is invariant under vector base changes. In (+) enantiomer case, the characteristic eigenvalue equation is

$$\det\|\chi_k - \lambda \mathbf{I}\| = 0 \qquad (16)$$

where $\lambda$ are the eigenvalues and $\mathbf{I}$ is the identity matrix.

The following characteristic polynomials can be derived

$$(1-\lambda)^4 = 0 \qquad \text{for } \chi_1 \qquad (17)$$

$$(1-\lambda)^2(1+\lambda+\lambda^2) = 0 \qquad \text{for } \chi_2, \chi_3, \chi_4, \chi_6, \chi_8, \chi_9, \chi_{10}, \chi_{12} \qquad (18)$$

$$(1-\lambda)^2(1+\lambda)^2 = 0 \qquad \text{for } \chi_5, \chi_7, \chi_{11} \qquad (19)$$

In the case of (−) enantiomer, we have

$$\det\|\bar{\chi}_k - \lambda \mathbf{I}\| = 0 \qquad (20)$$

and the characteristic polynomials are

$$(1-\lambda)^3(1+\lambda) = 0 \qquad \text{for } \bar{\chi}_1, \bar{\chi}_2, \bar{\chi}_3, \bar{\chi}_6, \bar{\chi}_8, \bar{\chi}_{11} \qquad (21)$$

$$(1-\lambda)(1+\lambda)(\lambda^2+1)=0 \qquad \text{for } \bar{\chi}_4, \bar{\chi}_5, \bar{\chi}_7, \bar{\chi}_9, \bar{\chi}_{10}, \bar{\chi}_{12} \qquad (22)$$

There are 6 independent eigenvalues:

$$\lambda_{1,2}=\pm 1 \qquad \lambda_{3,4}=\pm i \qquad \lambda_{5,6}=\frac{-1\pm i\sqrt{3}}{2} \qquad (23)$$

Inserting them into Eqs. [16] and [20], it is easy to determine the eigenvectors

$$(\chi_k - \lambda \mathbf{I})\mathcal{M} = 0, \qquad (\bar{\chi}_k - \lambda \mathbf{I})\mathcal{M} = 0 \qquad (24)$$

with obvious calculations depending on the choice of $\chi_k$ and $\bar{\chi}_k$. $\mathcal{M}$ is given by Eq. [3]. It is worth noting that the number of independent eigenvalues (and then eigenvectors) is related to the number of independent elements in each member of the group $O(N)$ we are considering. $N^2$ is the total number of elements, while $\frac{1}{2}N(N+1)$ are the orthogonality conditions, so we have

$$N^2 - \frac{1}{2}N(N+1) = \frac{1}{2}N(N-1) \qquad (25)$$

For $O(4)$, it is 6, which is the number of independent generators of the group[11]. Such a number gives the "dimension" of the group.

## GENERALIZATION TO MOLECULES WITH N STEREOGENIC CENTERS

The results of the previous section can be extended to more general cases.

For a molecule with $n$ stereogenic centers, we can define $n$ planes of projection. The bonds among the centers have to be taken into account. If a molecule with one center has four bonds, a molecule with two centers has seven bonds and so on. The general rule is

$$n = \text{centers} \quad \Leftrightarrow \quad 4n - (n-1) = 3n+1 \text{ bonds} \qquad (26)$$

assuming simply connected tetrahedrons. When atoms act as "spacers" between tetrahedral chiral centers, the number of bonds changes from $3n+1$ to $4n$. A molecule with $n$ stereogenic centers, starting from Eq. [2], is then given by the sum vector

$$\mathcal{M}_n = \sum_{k=1}^{n} \sum_{j=1}^{3n+1} \rho_{jk} e^{i\theta_{jk}} \tag{27}$$

where $k$ is the "center-index" and $j$ is the "bond-index". A projective plane of symmetry is fixed for any $k$. The couple $\{\rho_{jk}, \theta_{jk}\} \equiv \{0,0\}$ defines the center in any plane. In other words, a molecule $\mathcal{M}_n$ is assigned by the two sets of numbers

$$\{\rho_{1k}, ...\rho_{jk}, ...\rho_{(3n+1)k}\}$$
$$\{\theta_{1k}, ...\theta_{jk}, ...\theta_{(3n+1)k}\} \tag{28}$$

Chirality emerges when two molecules with identical structural formulae, as shown algebraically in the previous section, are not superimposable. After a rotation, two bonds result superimposable, while the other two are inverted.

On the contrary, if after rotation and superimposition, the molecules are identical, the situation is achiral. Such a treatment can be repeated for any projective symmetry plane which can be defined for the $n$ centers. The possible results are that the molecule is fully invariant after rotation(s) and superimposition with respect to its mirror image (achiral); the molecule is partially invariant after rotation(s) and superimposition, i.e. some tetrahedrons are superimposable while others are not (diastereoisomers); the molecule presents an inversion for each stereogenic center (enantiomers).

Also for these general cases, the considerations of previous section can be applied. Let us write Eq. [27] as

$$\mathcal{M}_n = \sum_{k=1}^{p} \overline{\mathcal{M}}_k + \sum_{k=p+1}^{n} \mathcal{M}_k \tag{29}$$

where $\overline{\mathcal{M}}_k$ and $\mathcal{M}_k$ are generic tetrahedrons on which are acting the operators $\overline{\chi}_l^k$ and $\chi_l^k$ respectively; $k$ is the center index running from 1 to $n$; $l$ is the operator index ranging from 1 to 12. For any tetrahedron, we can have the two possibilities

$$\mathcal{M}_k = \chi_l^k \mathcal{M}_k^{(0)}, \qquad \overline{\mathcal{M}}_k = \overline{\chi}_l^k \mathcal{M}_k^{(0)} \qquad (30)$$

where $\mathcal{M}_k^{(0)}$ is the starting fundamental representation of the *k*-tetrahedron given by the column vector in Eq. [3]. Explicitly we have

$$\mathcal{M}_k^{(0)} = \begin{pmatrix} \Psi_{1k} \\ \Psi_{2k} \\ \Psi_{3k} \\ \Psi_{4k} \end{pmatrix} \qquad (31)$$

In other words, $\mathcal{M}_k$ and $\overline{\mathcal{M}}_k$ are the result of the application of one of the above matrix operators on the starting column vector $\mathcal{M}_k^{(0)}$.

A particular discussion deserves the index *p*. Its range is

$$0 \leq p \leq n \qquad (32)$$

It is the number of permutations (at most one for any center) which occur when the operators $\overline{\chi}_l^k$ act on tetrahedrons. It corresponds to the number of reflections occuring in a *n*-center tetrahedral chain. No inversions, but rotations occur when $\chi_l^k$ operators act on the molecule. With this rule in mind, it follows that

$$\mathcal{M}_n = \sum_{k=1}^{n} \mathcal{M}_k, \qquad p = 0 \qquad (33)$$

is an achiral molecule (in this case only $\chi_l^k$ operators act on $\mathcal{M}_k^{(0)}$);

$$\mathcal{M}_n = \sum_{k=1}^{p} \overline{\mathcal{M}}_k + \sum_{k=p+1}^{n} \mathcal{M}_k, \qquad 0 < p < n \qquad (34)$$

is a diastereoisomer since $[n-(p+1)]$ tetrahedrons result superimposable after rotations, while *p*-ones are not superimposable, having, each of them, undergone an inversion of two of their bonds. An enantiomer results if

$$\mathcal{M}_n = \sum_{k=1}^{n} \overline{\mathcal{M}}_k, \qquad n = p \qquad (35)$$

where every tetrahedron results a mirror image of its starting situation after the application of any of the $\bar{\chi}_i^k$ operators.

The chirality selection rule, deduced in a previous paper,[8] is fully recovered. Let us recall it for the sake of completeness: central chirality is assigned by the number $\chi$ given by the couple $n, p$ that is

$$\chi = \{n, p\} \qquad (35)$$

where we define

$$\chi = \text{chirality index}$$

$$n = \text{principal chiral number}$$

$$p = \text{secondary chiral number}$$

$n$ is the number of stereogenic centers, $p$ is the number of inversions.

The sequence between achiral and chiral molecules is given by

$$\chi \equiv \{n, 0\} \quad \textit{achiral molecules}$$

$$\chi \equiv \{n, p < n\} \quad \textit{diastereoisomers}$$

$$\chi \equiv \{n, n\} \quad \textit{enantiomers}$$

In other words, this selection rule gives a classification of tetrahedral molecular chains by their chirality structure.

## SUMMARY

In this paper, the algebraic structure of central molecular chirality, starting from the well-known empirical Fischer rules, was deduced. The operators of such an algebra are obtained from the 24 projections which a single tetrahedron can generate in $S$ and $R$ configurations. They constitute a matrix representation of $O(4)$ orthogonal group of dimension 6. Twelve of them are rotations, while the other twelve are inversions, as can be derived from the value of their determinant, which is 1 for rotations and −1 for inversions. It is useful to remember that all the projections are algebraically generated starting from a fundamental one where the positions of chemical groups are established *a priori* in a clockwise sequence 1→2→3→4.

The generalization to chains of tetrahedrons is straightforward leading to the distinction among achiral molecules, diastereoisomers and enantiomers. These classes of molecules can be parametrized by the chirality index $\chi = \{n, p\}$, geometrically deduced in a previous report.[8] The number *n* enumerates stereogenic centers; *p* indicates the inversions occuring when some of the tetrahedrons of a given molecule cannot be superimposed to their mirror image after rotation.

Finally, the rule holds only for simply connected chains of tetrahedrons (and eventually when "spacer" atoms are placed between bonds).

**LITERATURE CITED**

1. Kelvin WT. Baltimore lectures on molecular dynamics and the wave theory of light. London: Clay C. J.; 1904:619.

2. a) Eliel EL, Wilen SH, Mander LN. Stereochemistry of organic compounds. New York: John Wiley & Sons; 1994. b) Eliel EL. Elements of stereochemistry. New York: John Wiley & Sons; 1969.

3. Ōki M. The chemistry of rotational isomers. New York: Springer; 1993.

4. Runge W. The chemistry of the allenes. vol. 2, Landor, S R: Academic Press; 1982.

5. Laarhoven WH, Prinsen WJC. Carbohelicenes and heterohelicenes. Top. Curr. Chem. 1984;125:63-130.

6. a) Sokolov VI. Introduction to theoretical stereochemistry. New York: Gordon and Breach Science: 1991. b) Dreiding AS, Wirth K. MATCH 1980;8:341-352. c) Tratch SS, Zefirov NS. Combinatorial models and Algorithmns in chemistry. Ladder of combinatorial objects and its application to formalization of structural problems of organic chemistry. In: Stepanov NF, editor. Principles of symmetry and systemology in chemistry. Moscow: Moscow State University Press; 1987. p 54. d) Johnson M, Tsai C-C, Nicholson V. J Math. Chem. 1991;7:3-38.

7. a) Ruch E, Schönhofer A. Näherungsformeln für spiegelungsantimetrische Moleküleigenschaften. Theor Chim Acta 1968;10:91-110. b) Ruch E, Schönhofer A. Theorie der Chiralitätsfunktionen. Theor Chim Acta 1970;19:225-287. c) Ruch E. Algebraic aspect of the chirality phenomenon in chemistry. Acc Chem Res 1972;5: 49-56. d) King RB. Chirality polynomials. J Math Chem 1988;2:89-115. e) King RB. Experimental test of chirality algebra. J Math Chem 1991;7:69-84. f) King RB. Chirality algebra. In: Mezey PG, editor. New developments in molecular chirality. Dordrecht: Kluwer; 1991. p 131-164. g) King RB. Chirality algebra and the right-left classification problem. In: Pályi G, Zucchi C, Caglioti L, editors. Advances in biochirality. Lausanne: Elsevier; 1999. p 13-34.

8. Capozziello S, Lattanzi A. Geometrical approach to central molecular chirality: a chirality selection rule. Chirality 2003 (in press).

9. Kaku M. Quantum field theory. Oxford: Oxford University Press; 1993.

10. Lang S. Linear algebra. Reading MA: Addison-Wesley; 1966.

11. Gilmore R. Lie groups, Lie algebras, and some of their applications. Malabar, Florida: Krieger Publishing Company; 1994.